\documentclass{icrc2009}

\usepackage{graphicx}   
\usepackage{caption}    
\usepackage[font=footnotesize]{subfig} 
\usepackage{fixltx2e}
\usepackage{url}

\newcommand{\shorttitle}[1]%
{\markboth{Proceedings of the 31\MakeLowercase{$^{st}$} ICRC, {\L}\'{o}d\'{z} 2009}{#1} }
\newcommand{\etal}{\MakeLowercase{\textit{et al. }}} 

\newcommand{\apj}{\textit{ApJ}} 
 
\newcommand{\aaps}{\textit{A}\&\textit{AS}} 
\newcommand{\mnras}{\textit{MNRAS}} 
\newcommand{\pasj}{\textit{PASJ}} 
\newcommand{\aap}{\textit{A}\&\textit{A}} 
\newcommand{\aj}{\textit{AJ}} 

\begin{document}
\title{Multiwavelength observation from radio through very-high-energy $\gamma$-ray of OJ~287 during the 12-year cycle flare in 2007 }

\author{\IEEEauthorblockN{Masaaki Hayashida\IEEEauthorrefmark{1}, Giacomo Bonnoli\IEEEauthorrefmark{2}, Antonio Stamerra\IEEEauthorrefmark{2}, }
\IEEEauthorblockN{Elina Lindfors\IEEEauthorrefmark{3}, Kari Nilsson\IEEEauthorrefmark{3} and Masahiro Teshima\IEEEauthorrefmark{4} for the MAGIC collaboration }
                      	\IEEEauthorblockN{and }
		   \IEEEauthorblockN{Hiromi Seta\IEEEauthorrefmark{5}, Naoki Isobe\IEEEauthorrefmark{6}, Makoto S. Tashiro\IEEEauthorrefmark{5},}
		   \IEEEauthorblockN{Koichiro Nakanishi\IEEEauthorrefmark{7}, Mahito Sasada\IEEEauthorrefmark{8}, Yoshito Shimajiri\IEEEauthorrefmark{7}\IEEEauthorrefmark{9} and Makoto Uemura\IEEEauthorrefmark{10}}
                            \\
\IEEEauthorblockA{\IEEEauthorrefmark{1}Kavli Institute for Particle Astrophysics and Cosmology, SLAC National Accelerator Laboratory, CA, 94025, USA}
\IEEEauthorblockA{\IEEEauthorrefmark{2}Universit\`a  di Siena, and INFN Pisa, I-53100 Siena, Italy}
\IEEEauthorblockA{\IEEEauthorrefmark{3}Tuorla Observatory, University of Turku, FI-21500 Piikki\"o, Finland}
\IEEEauthorblockA{\IEEEauthorrefmark{4}Max-Planck-Institut f\"ur Physik, D-80805 M\"unchen, Germany}
\IEEEauthorblockA{\IEEEauthorrefmark{5}Department of Physics, Saitama University, Saitama 338-8570, Japan}
\IEEEauthorblockA{\IEEEauthorrefmark{6}Department of Astronomy, Kyoto University, Kyoto 606-8502, Japan}
\IEEEauthorblockA{\IEEEauthorrefmark{7}Nobeyama Radio Observatory, Nagano 384-1305, Japan}
\IEEEauthorblockA{\IEEEauthorrefmark{8}Department of Physical Science, Hiroshima University, Higashi-Hiroshima 739-8526, Japan}
\IEEEauthorblockA{\IEEEauthorrefmark{9}Department of Astronomy School of Science, University of Tokyo, Tokyo 113-0033, Japan}
\IEEEauthorblockA{\IEEEauthorrefmark{10}Astrophysical Science Center, Hiroshima University, Higashi-Hiroshima 739-8526, Japan}
}

\shorttitle{Hayashida, M, \etal OJ~287 multiwavelength observations in 2007}
\maketitle

\begin{abstract}
We performed simultaneous multiwavelength observations of OJ~287 with the Nobeyama Millimeter Array for radio, 
the KANATA telescope and the KVA telescope for optical, the Suzaku satellite for X-ray and the MAGIC telescope for very high energy (VHE) $\gamma$-ray in 2007.
The observations were conducted for a quiescent state in April and in a flaring state in November-December.
We clearly observed increase of fluxes from radio to X-ray bands during the flaring state while MAGIC could not detect significant VHE $\gamma$-ray emission from the source. We could derive an upper limit (95\% confidence level) of 1.7\% of the Crab Nebula flux above 150 GeV from about 41.2 hours of the MAGIC observation. A simple SSC model suggests that the observed flaring activity could be caused by evolutions in the distribution of the electron population rather than changes of the magnetic field strength or Doppler beaming factor in the jet.
\end{abstract}


 \section{Introduction}
OJ~287 ($z = 0.306$ \cite{Stickel}) is one of the archetypal and most studied blazars. 
An outstanding characteristic of the object is 
its recurrent optical outbursts with a period of $11.65$ years, 
as revealed by optical data 
spanning more than $100$ years \cite{SMBH}. ``The OJ~$94$ project'' \cite{OJ94} confirmed the periodicity and revealed
that the optical outbursts consist of two peaks corresponding to flares 
with an interval of about one year \cite{double_peak}. 
OJ~287 is suggested to be a binary black hole system in which a secondary
black hole pierces the accretion disk of the primary black hole and
produces two impact flashes per period \cite{Valtonen_nature}. The differences between the two flares may be interpreted as following;
the first flare have a thermal origin in the vicinity of the black hole and the accretion disk while the second one originate from synchrotron radiation from the jet \cite{disk-jet}.
However, we have not yet obtained any convincing evidence supporting this interpretation. 

The multiwavelength spectral energy distribution (SED) has 
the potential to resolve the physical state of OJ~287 during the flares.
In general, the SED of blazars is characterized 
by two broad humps~(e.g., \cite{Fos98}); 
the low-energy component, 
with wavelengths in the range between radio to ultraviolet and X-ray, 
is widely regarded as synchrotron radiation from relativistic electrons within the jet.
The high-energy component, with wavelengths in the range between X-rays and $\gamma$-rays, 
is interpreted as inverse-Compton (IC) scattering. 
In one of the simple emission models, named 
``synchrotron self-Compton (SSC) model'', 
relativistic electrons scatter synchrotron photons produced by the same population of electrons (e.g., \cite{Ghi98}).
For low-frequency peaked BL Lac objects (LBLs), a class of blazars to
which OJ~287 belongs, 
the synchrotron peak is located in the range between sub-mm and optical wavelengths~\cite{Pad95}.
IC scattering in LBLs can emit radiation up to very-high-energy (VHE: E$>$50 GeV) $\gamma$-rays during their optical high states; 
VHE $\gamma$-ray emission has been detected, for example, from BL Lacertae~\cite{BLLac} and S5~0716+714~\cite{Tes08}.
These two components usually intersect with each other in the X-ray band (e.g., \cite{S50716}).

In the period between $2005$ and $2008$,
OJ~287 was predicted to move to the last 
active phase, 
and was in fact reported to exhibit the first optical outburst 
in $2005$ November \cite{Valtonen_nature,OJ287_2005_2}.
Since the second flare of the source was expected to be in the fall of $2007$ 
(\cite{Valtonen2,Kidger}),
we organized two simultaneous multiwavelength observation campaigns from radio through VHE $\gamma$-ray, 
in the quiescent state in April 2007 (MWL~I) between the two outbursts 
and in the second flaring state in November-December 2007 (MWL~II), with the
objective to reveal the characteristics of the second flare,
in comparison with the quiescent state.

In this contribution, we present the results of the multiwavelength campaigns with detailed observation results of the VHE $\gamma$-ray band by MAGIC,  and discuss the overall SEDs using a SSC model.
More extensive discussions of the campaigns including detailed results of the Suzaku X-ray satellite and other wavelength observations can be found in~\cite{Seta}.

\section{Multiwavelength Observations and Results}

\subsection{VHE $\gamma$-ray band by MAGIC}
We used the MAGIC telescope to search for VHE $\gamma$-rays 
emission from OJ~287 during the both MWL campaigns.
MAGIC is a single dish Imaging Atmospheric Cherenkov Telescope with a 17-m diameter main reflector. 
The telescope is located in the Canary Island of La Palma, in regular operation since 2004
with a low energy threshold of $50$ -- $60$ GeV 
(trigger threshold at small zenith angles;~\cite{MAGICCrab}). 

In MWL~I, MAGIC observed the source during $3$ nights. 
The zenith angle of the observations ranges from $8^{\circ}$ to
$29^{\circ}$. The observations were performed in so-called ON-OFF
observation mode. The telescope was pointing directly to the source,
recording ON-data. The background was estimated from
additional observations of regions where no $\gamma$-ray is expected,
OFF-data, which were taken with sky conditions similar to ON-data. 
Data runs with anomalous trigger rates due to bad observation conditions
were rejected from the analysis. 
The remaining data correspond to $4.5$ hours of ON and $6.5$ hours of OFF data.
In November and December $2007$ for MWL~II,
MAGIC observed
in a zenith angle range from $8^{\circ}$ to $31^{\circ}$ in the 
``wobble mode''~\cite{wobble}, where the object was observed at
$0.4^{\circ}$ offset from the camera center. 
In this observation mode,
an ON-data sample and OFF-data samples can be extracted from the same observation run; 
in our case, we used 3 OFF regions to estimate the background.
In total, the data were taken during $22$ 
nights.
$41.2$ hours of data from $19$ nights passed the quality selection to be
used for further analysis. 

The VHE $\gamma$-ray data taken for MWLs~I and II were analyzed using the MAGIC
standard calibration and analysis software. 
Detailed information about the analysis chain is found
in~\cite{MAGICCrab}.  
In February $2007$, the signal digitization of MAGIC was upgraded
to $2$ $\rm{GSamples~s^{-1}}$ FADCs, and timing information is used to
suppress the 
contamination of light of the night sky and to obtain new shower image
parameters~\cite{MAGICtime} in addition to conventional \textit{Hillas} image
parameters~\cite{Hillas}. 

These parameters were used for $\gamma$/hadron separation by means of
the ``Random Forest (RF)'' method~\cite{MAGICRF}. 
The $\gamma$/hadron separation based on the RF method was tuned to give a $\gamma$-cut efficiency of $70~\%$. 
Finally, the $\gamma$-ray signal was determined by comparing between ON and normalized-OFF data 
in the |ALPHA| parameter\footnote{the angle between the shower image principal axis
and the line connecting the image center of gravity with the camera center.} 
distribution, in which the $\gamma$-ray signal should show up as an excess at small values. 
Our analysis requires a $\gamma$-cut efficiency of $80~\%$ for the final |ALPHA| selection.
The energy of the $\gamma$-ray events are also estimated using the RF method.

A search of VHE $\gamma$-rays from OJ~287 was performed with data
taken for MWLs~I and II in three distinct energy bins.
No significant excess was found in any data samples. 
Upper limits with $95~\%$ confidence level in the number of excess events
were calculated using the method of \cite{Rolke}, taking into
account a systematic error of 30~\%. The number of excess events was converted
into flux upper limits assuming a photon index of $- 2.6$, 
corresponding to the value used in our Monte-Carlo samples of $\gamma$-rays.
The derived upper limits in the three energy bins for each period are summarized in table~\ref{table:UL_log}. 

\begin{figure}[ht]
 \centering
  \includegraphics[width=2.6in]{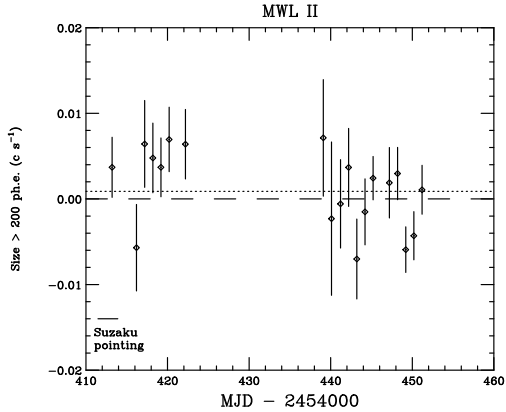}
\caption{
Excess event rate with SIZE above 200 photoelectrons 
(with a corresponding energy threshold of 150~GeV), 
observed with the MAGIC telescope in MWL~II.
The dotted line indicates the average count rate. 
}
\label{fig:magic_LC}
\end{figure}

A search for VHE flares with a short-time scale was also performed with the data taken for MWL~II. 
Figure~\ref{fig:magic_LC} shows the nightly count rate of the excess
events after all cuts including a SIZE cut above 200 photoelectrons,
corresponding to an energy threshold of 150 GeV. Fitting 
a constant to the observed flux yields 
$\chi^2/{\rm d.o.f.} = 25.55/18$ (a probability of $11~\%$), and thus indicating
no evidence of a VHE flare during
this period.

\begin{table}[ht]
\caption{Results of the search for VHE $\gamma$-ray emissions from OJ~287. }

\label{table:UL_log}  
\begin{tabular}{lccc}
\hline \hline
MWL~I & & &\\
\hline 
Threshed Energy\footnotemark[1]    & $80$          & $145$       & $310$\\
ON events\footnotemark[2]        & $40056$       & $1219$      & $42$\\
OFF events\footnotemark[3]        & $40397\pm226$ & $1340\pm38$ & $39.5\pm6.3$\\
significance ($\sigma$)\footnotemark[4] & $-1.13$       & $-0.94$     & $-0.47$\\
U.L. of excess\footnotemark[5]   & $394$         & $75.1$      & $21.9$\\
Flux$_{95\% \rm{U.L.}}$\footnotemark[6] 
                                           & $59.8$        & $11.1$      & $2.83$\\
Crab Flux(\%)\footnotemark[7]          &   $8.5$       &  $3.3$      &  $2.4$ \\
\hline  \hline 
MWL~II  & & &\\
\hline 
Threshed Energy\footnotemark[1]    & $85$          & $150$       & $325$\\
ON events\footnotemark[2]         & $281885$      & $12582$     & $578$\\
OFF events\footnotemark[3]        & $282342\pm493$& $12573\pm65$& $576\pm14$ \\
significance ($\sigma$)\footnotemark[4] & $-0.75$       & $0.07$      & $0.07$\\
U.L. of excess\footnotemark[5]   & $1218$   & $330$       & $71.6$\\
Flux$_{95\% \rm{U.L.}}$\footnotemark[6] 
                                           & $22.1$        & $5.64$      & $1.18$\\
Crab Flux(\%)\footnotemark[7]       &  $3.4$        &     $1.7$   &  $1.1$\\
\hline
  \end{tabular}
\footnotemark[1] in [GeV]. Corresponding to peak energies of $\gamma$-ray MC samples after all cuts.
\footnotemark[2] Number of measured ON events.
\footnotemark[3] Normalized number of OFF events and related error.
\footnotemark[4] Based on equation~(17) in~\cite{LiMa}.
\footnotemark[5] 95\% upper limit of the number of excess events with 30\% systematic error.
\footnotemark[6] in [$\times10^{-12}$ cm$^{-2}$ s$^{-1}$]. Flux upper limit assuming a photon index of $-2.6$ for the calculation of the effective area.
\footnotemark[7] Corresponding Crab flux in each energy range based on measurements of the Crab pulsar performed with the MAGIC telescope ~\cite{MAGICCrab}.
\end{table}

\subsection{Other energy bands}
In the X-ray band, the Suzaku satellite~\cite{Mituda} observed OJ~287 in the quiescent state (MWL~I) between 19:47:00 UT 2007 April 10 and 11:10:19 UT April 13, 
and the second flare (MWL~II) between 11:24:00 UT 2007 November 7 and 21:30:23 UT November 9. 
Significant X-ray signals were detected in the $0.5$ -- $10$ keV range in both observations.
Hard X-ray signals in $12$ -- $27$~keV was also clearly detected with a significance of $5.0~\sigma$ in MWL~II while those signals were not significant in MWL~I. 

Optical flux was monitored by the KANATA telescope in Hiroshima, Japan (in $V$, $J$, and $K_s$-bands), and the KVA telescope in the Canary Island of La Palma (in $R$-band).
Radio continuum emission from OJ~287 at 86.75~GHz and 98.75~GHz was also observed with the Nobeyama Millimeter Array (NMA) in Nobeyama, Japan.

\begin{figure}[ht]
 \centering
  \includegraphics[width=3.2in]{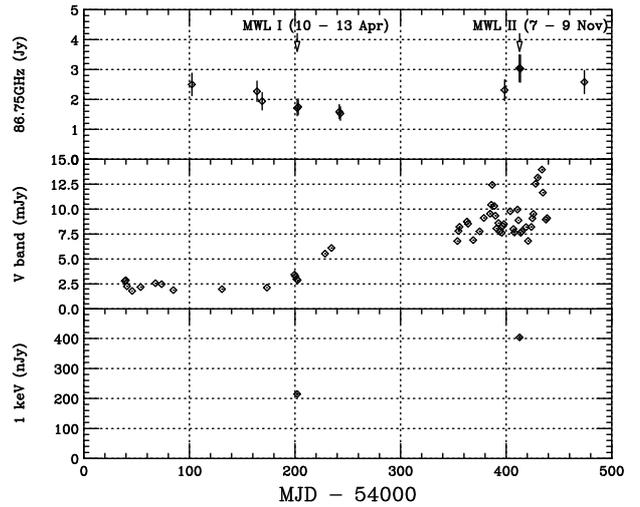}
\caption{
The multiwavelength lightcurves of OJ~287.
The top panel shows the radio flux at $86.75$~GHz as observed with NMA, while
the middle panel shows the optical flux in the $V$-band as observed with KANATA.
The radio and optical fluxes are averaged over each night.
The bottom panel shows the X-ray flux density at 1~keV.
Arrows indicate the Suzaku pointings in MWL~I and MWL~II.
}  
\label{fig:multi_lc}
\end{figure}

Figure~\ref{fig:multi_lc} summarizes the multiwavelength lightcurves of the radio, optical ($V$) and X-ray bands obtained 
between September 2006 and January 2008.
While the optical flux of OJ~287 was below $3~\rm{mJy}$ in the $V$-band before MWL~I,  
the brightness of the source started increasing after MWL~I to become the flaring state ($> 7$ mJy) in September $2007$.
The optical data show a monotonous decrease in a time scale of $\sim$ 4 days during MWL~II by a factor of 1.3.
The radio flux was generally higher during the optical flare in MWL~II than in MWL~I. The X-ray flux in MWL~II also increased by a factor of 2 compared to the flux in MWL~I.
Table~\ref{table:radio_opt_log} shows the flux of each energy bands during the Suzaku pointing in both MWLs~I and II.

 \begin{table*}[th]
\caption{Radio, optical and X-ray fluxes obtained during the Suzaku pointing in MWL~I and MWL~II.}
\label{table:radio_opt_log}
  \centering
  \begin{tabular}{lcccccccc}
  \hline \hline
    obs & \multicolumn{2}{c}{radio flux (Jy)}  & \multicolumn{4}{c}{optical flux (mJy)} & \multicolumn{2}{c}{X-ray flux (nJy)}  \\ 
      & 86.95~GHz\footnotemark[1] & 98.75~GHz\footnotemark[1] & $K_{s}$ \footnotemark[2]& $J$ \footnotemark[2]& $R$ \footnotemark[3]& $V$\footnotemark[2]  &   $S_{\rm 1keV}$\footnotemark[4] & $\Gamma$\footnotemark[5] \\
\hline 
MWL~I  & $1.73 \pm 0.26$ & $1.75 \pm 0.26$ & $17.74 \pm 0.33$        & $8.82 \pm 0.03$  & $3.20 \pm 0.05$ & $3.03 \pm 0.01$ &  $215 \pm 5$ & $1.65\pm0.02$ \\
MWL~II & $3.04 \pm 0.46$ & $2.98 \pm 0.46$ & $55.95^{+7.69}_{-6.76}$ & $27.02 \pm 0.21$ & $8.70\pm0.14$   & $8.93 \pm 0.05$  &  $404^{+6}_{-5}$ & $1.50\pm0.01$  \\
  \hline
    \end{tabular}
    
 \footnotemark[1] NMA data.
\footnotemark[2] KANATA data.
\footnotemark[3] KVA data.
\footnotemark[4] flux density at 1 keV of Suzaku data.
\footnotemark[5] photon index of the Suzaku data (0.5-10 keV).
  \end{table*}

\section{Discussion}

\begin{figure}[ht]
 \centering
  \includegraphics[width=3.2in]{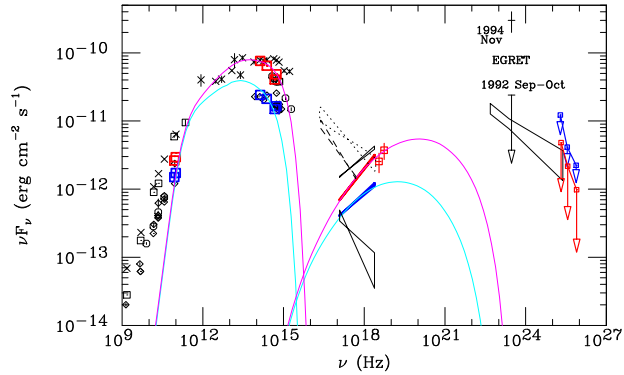}
\caption{
The SED of OJ~287 during the Suzaku observations in MWL~I (blue) and MWL~II (red). 
The radio and optical data are shown with squares and,
the X-ray data are shown with bow ties.  
The upper limit of the VHE $\gamma$-ray spectrum are measured values, 
shown with downward arrows.
The light blue lines and the purple lines indicate 
the simple one-zone SSC model for MWL~I and MWL~II, respectively. 
The black data points show radio, optical, and $\gamma$-ray data 
from non-simultaneous observations. 
The X-ray spectra with EXOSAT, ROSAT, and ASCA are 
drawn with dotted, dashed, and solid lines, respectively 
(\cite{Idesawa,Isobe} and references therein).
The $\gamma$-ray spectrum obtained with Fermi during the first 3 month 
observation (August -- October 2008) is shown with a bow tie~\cite{fermi_detect}. The $\gamma$-ray emission can be attributed to the external Compton radiation.
}
\label{fig:SED}
\end{figure}

Figure~\ref{fig:SED} shows the overall SED of OJ~287 obtained during the MWL~I and MWL~II, as well as some historical data.
The low frequency synchrotron component has a spectral turnover at around $5 \times 10^{14}$ Hz.
The X-ray spectrum exceeds the extrapolation
from the optical synchrotron spectra in both observations. 
Therefore, we naturally attribute the observed X-ray spectra 
to the IC component rising toward the higher frequency range.
The SED indicates that both the synchrotron and IC intensities increased from MWL~I to MWL~II without any significant shift of the synchrotron peak frequency.

As a working hypothesis, here we assume simply that
the variation 
of the SED was caused by a change 
in electron energy density (or number density) 
and/or the maximum Lorentz factor of the electrons,
with stable magnetic field, volume of emission region, 
minimum Lorentz factor, and break of electron energy distribution (e.g., \cite{Mrk421}).
In order to evaluate this hypothesis, 
we applied a one-zone SSC model to the SED
by using the numerical code developed by \cite{Kataoka}.  
The electron number density spectrum was assumed to be a broken-power law and  
the index of the electron spectrum ($p$) below the break Lorentz factor was
determined by the X-ray photon index as $p=2\Gamma - 1 = 2.3$ and $2.0$, 
in MWL~I and MWL~II, respectively.  
We obtained the following seven free parameters to describe the observed SED: 
the Doppler factor ($\delta$), the electron energy density ($u_{\rm e}$), 
the magnetic field ($B$), the blob radius ($R$), 
and the minimum, break, and maximum Lorentz factor of the electrons 
($\gamma_{\rm min}$, $\gamma_{\rm break}$, and $\gamma_{\rm max}$, respectively).
Adopting the optical variability time scale ($T_{\rm var} \sim$ 4 days) in MWL~II, 
the relation between $\delta$ and $R$ should be subjected to 
$R < c T_{\rm var} \delta / (1+z) = 1.2 \times 10^{17} (T_{\rm var} / 4{\rm days}) (\delta / 15)~{\rm [cm]}$
where $c$ and $z$ are the speed of light and the redshift of the source,
respectively.
We could reproduce SEDs in both MWL~I and II by the SSC model as parameters are summarized in table~\ref{table:SSC_log}.
The differences between two states can be found in $p$, $u_{\rm e}$ and $\gamma_{\rm {max}}$ while the other parameters remained unchanged.
Thus, we conclude that  the increase in the electron energy density produced the second flare.

In 2008, Fermi Gamma-ray Space Telescope successfully detected a $\gamma$-ray spectrum from the quiescent state of OJ~287 during
its first three months \cite{fermi_detect}, as shown in figure~\ref{fig:SED}.
The measured $\gamma$-ray flux significantly exceeds our simple SSC model flux.
This may indicate a contribution of external Compton radiation to the $\gamma$-ray emission from OJ~287. 
A simultaneous multiwavelength observation with Fermi will be essential to test emission models with the external Compton radiation to the $\gamma$-ray component.

\begin{table}[ht]
\caption{Physical parameters for the SSC model.}
\label{table:SSC_log}
  \centering
\begin{tabular}{lcc}
\hline \hline 
parameters\footnotemark[1] & MWL~I & MWL~II \\                 
\hline 
\multicolumn{1}{l}{$\delta$}              & \multicolumn{2}{c}{$15$}\\
\multicolumn{1}{l}{$R$ ($\rm{cm}$)}       & \multicolumn{2}{c}{$7.0 \times 10^{16}$}\\
\multicolumn{1}{l}{$B$ ($\rm{Gauss}$)}    & \multicolumn{2}{c}{$0.71$}\\
\multicolumn{1}{l}{$\gamma_{\rm{min}}$}   & \multicolumn{2}{c}{$70$}\\
\multicolumn{1}{l}{$\gamma_{\rm{break}}$} & \multicolumn{2}{c}{$700$}\\
$\gamma_{\rm{max}}$   & $3300$               & $4500$ \\
$p$                   & $2.3$                & $2.0$ \\
$u_{\rm m}$ ($\rm{erg~cm^{-3}}$) & $2.0 \times 10^{-2}$ & $2.0 \times 10^{-2}$ \\
$u_{\rm e}$ ($\rm{erg~cm^{-3}}$) & $1.5 \times 10^{-3}$ & $2.1 \times 10^{-3}$ \\ 
\hline
  \end{tabular}
  \par\noindent
\footnotemark[1] Notations are described in the text.\\
\end{table}


\section*{Acknowledgment} 
We would like to thank the Instituto de Astrofisica de
Canarias for the excellent working conditions at the
Observatorio del Roque de los Muchachos in La Palma. 
The support of the German BMBF and MPG, the Italian INFN 
and Spanish MICINN is gratefully acknowledged. 
This work was also supported by ETH Research Grant 
TH 34/043, by the Polish MNiSzW Grant N N203 390834, 
and by the YIP of the Helmholtz Gemeinschaft.
We also thank all members of the Suzaku team
for performing successful operation and calibration.
The Nobeyama Radio Observatory is a branch of the National Astronomical
Observatory of Japan, the National Institutes of Natural Sciences (NINS).
IRAF is distributed by the National Optical 
Astronomy Observatories, which are operated by the Association of 
Universities for Research in Astronomy, Inc., under a cooperative 
agreement with the National Science Foundation.


\begin{thebibliography}{99}
 
\bibitem{fermi_detect}
        Abdo, A.~A., et al. 2009, arXiv:0902.1559 
\bibitem{BLLac}
	Albert, J., et al. 2007, \apj, 666, L17
\bibitem{MAGICCrab}
	Albert, J., et al. 2008a, \apj, 674, 1037
\bibitem{MAGICRF}
	Albert, J., et al. 2008b, Nucl. Instrum. and Meth., A588, 424
\bibitem{MAGICtime} 
	Aliu, E., et al.\ 2009, Astroparticle Physics, 30, 293 
\bibitem{Fos98}
        Fossati, G., et al., \mnras, 299, 433	
\bibitem{wobble}
	Fomin, V. P., et al.,\ 1994, Astropart. Phys., 2, 137 
\bibitem{Ghi98}
        Ghisellini, G., et al.,\ 1998, \mnras, 301, 451
\bibitem{S50716}
	Giommi, P., et al., 1999, \aap, 351, 59
\bibitem{Hillas}
	Hillas, A.~M. 1985, Proc. 29th Int. Cosmic Ray Conf. (La Jolla), 3, 445
\bibitem{Idesawa} 
	Idesawa, E., et al.\ 1997, \pasj, 49, 631
\bibitem{Isobe} 
	Isobe, N., Tashiro, M., Sugiho, M., \& Makishima, K.\ 2001, \pasj, 53, 79 
\bibitem{Kataoka}
	Kataoka, J., Ph.D thesis, 2000, Univ. of Tokyo
\bibitem{Kidger} 
	Kidger, M.~R.\ 2000, \aj, 119, 2053 
\bibitem{LiMa}
	Li, T.-P., \& Ma, Y.-Q. 1983, \apj, 272, 317
\bibitem{Mituda} 
	Mitsuda, K., et al.\ 2007, \pasj, 59, 1 
\bibitem{Pad95}
        Padovani, P. \& Giommi, P. 1995, \apj, 444, 567	
\bibitem{Pursimo}
        Pursimo, T., et al.\ 2000, \aaps, 146, 141 
\bibitem{Rolke}
	Rolke, W. A., L{\'o}pez, A. M., \& Conrad, J. 2005, Nucl. Instrum. and Meth., A551, 493
\bibitem{OJ287_EGRET}
	Shrader, C.R., Hartman, R.C., Webb, J.R.,
	1996, \aap, 120, 599,
\bibitem{Seta} 
	Seta, H., et al.\ 2009, \pasj, submitted 
\bibitem{Stickel} 
	Stickel, M., Fried, J.~W., \& Kuehr, H.\ 1989, \aaps, 80, 103 
\bibitem{SMBH} 
	Sillanp\"{a}\"{a}, A., et al.\ 1988, \apj, 325, 628 
\bibitem{double_peak}
	Sillanp\"{a}\"{a}, A., et al.\ 1996, \aap, 315, L13 
\bibitem{OJ94} 
	Sillanp\"{a}\"{a}, A., et al.\ 1996, \aap, 305, L17 
\bibitem{Mrk421} 
	Takahashi, T., et al. 2000, \apj, 542, L105 
\bibitem{Tes08} 
	Teshima, M., et al.\ 2008, The Astronomer's Telegram, \#1500
\bibitem{Valtonen2} 
	Valtonen, M.~J., et al.\ 2006, \apj, 643, L9 
\bibitem{Valtonen_nature}
	Valtonen, M., et al., 2008, Nature, 452, 7189, 851
\bibitem{OJ287_2005_2}
	Valtonen, M., et al.,\ 2008, \aap, 477, 407
\bibitem{disk-jet} 
	Valtaoja, E., et al., \ 2000, \apj, 531, 744  
\bibitem{MAGICMC}
	Majumdar, P., et al.\ 2005, in Proc. 29th Int. Cosmic Ray Conf. (Pune, India), 5, 203
\end{thebibliography}
\end{document}